\documentclass[aps,prl,twocolumn,superscriptaddress,notitlepage]{revtex4-1}
\usepackage{graphicx} 
\usepackage{epsfig}
\usepackage{epstopdf}
\usepackage{amsmath}
\usepackage{amsfonts}
\usepackage{amssymb} 
\usepackage{color}
\usepackage{multirow}

\begin{document}

\title{Rosenbluth separation of the $\pi^0$ Electroproduction Cross Section off the Neutron}
\author{M.~Mazouz}
\email{mazouz@jlab.org}
\affiliation{Facult\'e des Sciences de Monastir, 5000 Tunisia}
\author{Z.~Ahmed}
\affiliation{Syracuse University, Syracuse, New York 13244, USA}
\author{H.~Albataineh}
\affiliation{Texas A\&M University-Kingsville, Kingsville, Texas 78363, USA}
\author{K.~Allada}
\affiliation{Massachusetts Institute of Technology,Cambridge, Massachusetts 02139, USA}
\author{K.~A.~Aniol}
\affiliation{California State University, Los Angeles, Los Angeles, California 90032, USA}
\author{V.~Bellini}
\affiliation{INFN/Sezione di Catania, 95125 Catania, Italy}
\author{M.~Benali}
\affiliation{Clermont universit\'{e}, universit\'{e} Blaise Pascal, CNRS/IN2P3, Laboratoire de physique corpusculaire, FR-63000 Clermont-Ferrand, France}
\author{W.~Boeglin}
\affiliation{Florida International University, Miami, Florida 33199, USA}
\author{P.~Bertin}
\affiliation{Clermont universit\'{e}, universit\'{e} Blaise Pascal, CNRS/IN2P3, Laboratoire de physique corpusculaire, FR-63000 Clermont-Ferrand, France}
\affiliation{Thomas Jefferson National Accelerator Facility, Newport News, Virginia 23606, USA}
\author{M.~Brossard}
\affiliation{Clermont universit\'{e}, universit\'{e} Blaise Pascal, CNRS/IN2P3, Laboratoire de physique corpusculaire, FR-63000 Clermont-Ferrand, France}
\author{A.~Camsonne}
\affiliation{Thomas Jefferson National Accelerator Facility, Newport News, Virginia 23606, USA}
\author{M.~Canan}
\affiliation{Old Dominion University, Norfolk, Virginia 23529, USA}
\author{S.~Chandavar}
\affiliation{Ohio University, Athens, Ohio 45701, USA}
\author{C.~Chen}
\affiliation{Hampton University, Hampton, Virginia 23668, USA}
\author{J.-P.~Chen}
\affiliation{Thomas Jefferson National Accelerator Facility, Newport News, Virginia 23606, USA}
\author{M.~Defurne}
\affiliation{Irfu, CEA, Universit\'{e} Paris-Saclay, 91191 Gif-sur-Yvette, France}
\author{C.W.~de~Jager}
\thanks{Deceased}
\affiliation{Thomas Jefferson National Accelerator Facility, Newport News, Virginia 23606, USA}
\thanks{Deceased}
\author{R.~de~Leo}
\affiliation{Universit\`{a} di Bari, 70121 Bari, Italy}
\author{C.~Desnault}
\affiliation{Institut de Physique Nucl\'eaire CNRS-IN2P3, Orsay, France}
\author{A.~Deur}
\affiliation{Thomas Jefferson National Accelerator Facility, Newport News, Virginia 23606, USA}
\author{L.~El~Fassi}
\affiliation{Rutgers, The State University of New Jersey, Piscataway, New Jersey 08854, USA}
\author{R.~Ent}
\affiliation{Thomas Jefferson National Accelerator Facility, Newport News, Virginia 23606, USA}
\author{D.~Flay}
\affiliation{Temple University, Philadelphia, Pennsylvania 19122, USA}
\author{M.~Friend}
\affiliation{Carnegie Mellon University, Pittsburgh, Pennsylvania 15213, USA}
\author{E.~Fuchey}
\affiliation{Clermont universit\'{e}, universit\'{e} Blaise Pascal, CNRS/IN2P3, Laboratoire de physique corpusculaire, FR-63000 Clermont-Ferrand, France}
\author{S.~Frullani}
\thanks{Deceased}
\affiliation{INFN/Sezione Sanit\`{a}, 00161 Roma, Italy}
\author{F.~Garibaldi}
\affiliation{INFN/Sezione Sanit\`{a}, 00161 Roma, Italy}
\author{D.~Gaskell}
\affiliation{Thomas Jefferson National Accelerator Facility, Newport News, Virginia 23606, USA}
\author{A.~Giusa}
\affiliation{INFN/Sezione di Catania, 95125 Catania, Italy}
\author{O.~Glamazdin}
\affiliation{Kharkov Institute of Physics and Technology, Kharkov 61108, Ukraine}
\author{S.~Golge}
\affiliation{North Carolina Central University, Durham, North Carolina 27701, USA}
\author{J.~Gomez}
\affiliation{Thomas Jefferson National Accelerator Facility, Newport News, Virginia 23606, USA}
\author{O.~Hansen}
\affiliation{Thomas Jefferson National Accelerator Facility, Newport News, Virginia 23606, USA}
\author{D.~Higinbotham}
\affiliation{Thomas Jefferson National Accelerator Facility, Newport News, Virginia 23606, USA}
\author{T.~Holmstrom}
\affiliation{Longwood University, Farmville, Virginia 23909, USA}
\author{T.~Horn}
\affiliation{The Catholic University of America, Washington, DC 20064, USA}
\author{J.~Huang}
\affiliation{Massachusetts Institute of Technology,Cambridge, Massachusetts 02139, USA}
\author{M.~Huang}
\affiliation{Duke University, Durham, North Carolina 27708, USA}
\author{G.M.~Huber}
\affiliation{University of Regina, Regina, Saskatchewan S4S 0A2, Canada}
\author{C.E.~Hyde}
\affiliation{Old Dominion University, Norfolk, Virginia 23529, USA}
\affiliation{Clermont universit\'{e}, universit\'{e} Blaise Pascal, CNRS/IN2P3, Laboratoire de physique corpusculaire, FR-63000 Clermont-Ferrand, France}
\author{S.~Iqbal}
\affiliation{California State University, Los Angeles, Los Angeles, California 90032, USA}
\author{F.~Itard}
\affiliation{Clermont universit\'{e}, universit\'{e} Blaise Pascal, CNRS/IN2P3, Laboratoire de physique corpusculaire, FR-63000 Clermont-Ferrand, France}
\author{Ho.~Kang}
\affiliation{Seoul National University, Seoul, South Korea}
\author{Hy.~Kang}
\affiliation{Seoul National University, Seoul, South Korea}
\author{A.~Kelleher}
\affiliation{College of William and Mary, Williamsburg, Virginia 23187, USA}
\author{C.~Keppel}
\affiliation{Thomas Jefferson National Accelerator Facility, Newport News, Virginia 23606, USA}
\author{S.~Koirala}
\affiliation{Old Dominion University, Norfolk, Virginia 23529, USA}
\author{I.~Korover}
\affiliation{Tel Aviv University, Tel Aviv 69978, Israel}
\author{J.J.~LeRose}
\affiliation{Thomas Jefferson National Accelerator Facility, Newport News, Virginia 23606, USA}
\author{R.~Lindgren}
\affiliation{University of Virginia, Charlottesville, Virginia 22904, USA}
\author{E.~Long}
\affiliation{Kent State University, Kent, Ohio 44242, USA}
\author{M.~Magne}
\affiliation{Clermont universit\'{e}, universit\'{e} Blaise Pascal, CNRS/IN2P3, Laboratoire de physique corpusculaire, FR-63000 Clermont-Ferrand, France}
\author{J.~Mammei}
\affiliation{University of Massachusetts, Amherst, Massachusetts 01003, USA}
\author{D.J.~Margaziotis}
\affiliation{California State University, Los Angeles, Los Angeles, California 90032, USA}
\author{P.~Markowitz}
\affiliation{Florida International University, Miami, Florida 33199, USA}
\author{A.~Mart\'i Jim\'enez-Arg\"uello}
\affiliation{Facultad de F\'isica, Universidad de Valencia, Valencia, Spain}
\affiliation{Institut de Physique Nucl\'eaire CNRS-IN2P3, Orsay, France}
\author{F.~Meddi}
\affiliation{INFN/Sezione Sanit\`{a}, 00161 Roma, Italy}
\author{D.~Meekins}
\affiliation{Thomas Jefferson National Accelerator Facility, Newport News, Virginia 23606, USA}
\author{R.~Michaels}
\affiliation{Thomas Jefferson National Accelerator Facility, Newport News, Virginia 23606, USA}
\author{M.~Mihovilovic}
\affiliation{University of Ljubljana, 1000 Ljubljana, Slovenia}
\author{N.~Muangma}
\affiliation{Massachusetts Institute of Technology,Cambridge, Massachusetts 02139, USA}
\author{C.~Mu\~noz~Camacho}
\affiliation{Clermont universit\'{e}, universit\'{e} Blaise Pascal, CNRS/IN2P3, Laboratoire de physique corpusculaire, FR-63000 Clermont-Ferrand, France}
\affiliation{Institut de Physique Nucl\'eaire CNRS-IN2P3, Orsay, France}
\author{P.~Nadel-Turonski}
\affiliation{Thomas Jefferson National Accelerator Facility, Newport News, Virginia 23606, USA}
\author{N.~Nuruzzaman}
\affiliation{Hampton University, Hampton, Virginia 23668, USA}
\author{R.~Paremuzyan}
\affiliation{Institut de Physique Nucl\'eaire CNRS-IN2P3, Orsay, France}
\author{A.~Puckett}
\affiliation{Los Alamos National Laboratory, Los Alamos, New Mexico 87545, USA}
\author{V.~Punjabi}
\affiliation{Norfolk State University, Norfolk, Virginia 23529, USA}
\author{Y.~Qiang}
\affiliation{Thomas Jefferson National Accelerator Facility, Newport News, Virginia 23606, USA}
\author{A.~Rakhman}
\affiliation{Syracuse University, Syracuse, New York 13244, USA}
\author{M.N.H.~Rashad}
\affiliation{Old Dominion University, Norfolk, Virginia 23529, USA}
\author{S.~Riordan}
\affiliation{Stony Brook University, Stony Brook, New York 11794, USA}
\author{J.~Roche}
\affiliation{Ohio University, Athens, Ohio 45701, USA}
\author{G.~Russo}
\affiliation{INFN/Sezione di Catania, 95125 Catania, Italy}
\author{F.~Sabati\'e}
\affiliation{Irfu, CEA, Universit\'{e} Paris-Saclay, 91191 Gif-sur-Yvette, France}
\author{K.~Saenboonruang}
\affiliation{University of Virginia, Charlottesville, Virginia 22904, USA}
\affiliation{Kasetsart University, Chatuchak, Bangkok, 10900, Thailand}
\author{A.~Saha}
\thanks{Deceased}
\affiliation{Thomas Jefferson National Accelerator Facility, Newport News, Virginia 23606, USA}
\author{B.~Sawatzky}
\affiliation{Thomas Jefferson National Accelerator Facility, Newport News, Virginia 23606, USA}
\affiliation{Temple University, Philadelphia, Pennsylvania 19122, USA}
\author{L.~Selvy}
\affiliation{Kent State University, Kent, Ohio 44242, USA}
\author{A.~Shahinyan}
\affiliation{Yerevan Physics Institute, Yerevan 375036, Armenia}
\author{S.~Sirca}
\affiliation{University of Ljubljana, 1000 Ljubljana, Slovenia}
\author{P.~Solvignon}
\thanks{Deceased}
\affiliation{Thomas Jefferson National Accelerator Facility, Newport News, Virginia 23606, USA}
\author{M.L.~Sperduto}
\affiliation{INFN/Sezione di Catania, 95125 Catania, Italy}
\author{R.~Subedi}
\affiliation{Georges Washington University, Washington, DC 20052, USA}
\author{V.~Sulkosky}
\affiliation{Massachusetts Institute of Technology,Cambridge, Massachusetts 02139, USA}
\author{C.~Sutera}
\affiliation{INFN/Sezione di Catania, 95125 Catania, Italy}
\author{W.A.~Tobias}
\affiliation{University of Virginia, Charlottesville, Virginia 22904, USA}
\author{G.M.~Urciuoli}
\affiliation{INFN/Sezione di Roma, 00185 Roma, Italy}
\author{D.~Wang}
\affiliation{University of Virginia, Charlottesville, Virginia 22904, USA}
\author{B.~Wojtsekhowski}
\affiliation{Thomas Jefferson National Accelerator Facility, Newport News, Virginia 23606, USA}
\author{H.~Yao}
\affiliation{Temple University, Philadelphia, Pennsylvania 19122, USA}
\author{Z.~Ye}
\affiliation{University of Virginia, Charlottesville, Virginia 22904, USA}
\author{L.~Zana}
\affiliation{Syracuse University, Syracuse, New York 13244, USA}
\author{X.~Zhan}
\affiliation{Argonne National Laboratory, Lemont, Illinois 60439, USA}
\author{J.~Zhang}
\affiliation{Thomas Jefferson National Accelerator Facility, Newport News, Virginia 23606, USA}
\author{B.~Zhao}
\affiliation{College of William and Mary, Williamsburg, Virginia 23187, USA}
\author{Z.~Zhao}
\affiliation{University of Virginia, Charlottesville, Virginia 22904, USA}
\author{X.~Zheng}
\affiliation{University of Virginia, Charlottesville, Virginia 22904, USA}
\author{P.~Zhu}
\affiliation{University of Virginia, Charlottesville, Virginia 22904, USA}
\collaboration{The Jefferson Lab Hall A Collaboration}

\date{\today}

\begin{abstract}
We report the first longitudinal/transverse separation of the deeply virtual exclusive $\pi^0$ electroproduction cross section off the neutron and coherent deuteron. The corresponding four structure functions $d\sigma_L/dt$, $d\sigma_T/dt$, $d\sigma_{LT}/dt$ and $d\sigma_{TT}/dt$ are extracted as a function of the momentum transfer to the recoil system at $Q^2$=1.75 GeV$^2$ and $x_B$=0.36. 
The $ed \to ed\pi^0$ cross sections are found compatible with the small values expected from theoretical models. The $en \to en\pi^0$ cross sections show a dominance from the response to transversely polarized photons,  and are in good agreement with calculations based on the transversity GPDs of the nucleon. By combining these results with previous measurements of $\pi^0$ electroproduction off the proton, we present a flavor decomposition of the $u$ and $d$ quark contributions to the cross section.
\end{abstract}


\maketitle

Understanding the internal three-dimensional structure of nucleons in terms of quarks and gluons is a major challenge of modern hadronic physics. Two complementary approaches have been used in the past in order to achieve this goal. On the one hand, nucleon form factors (FFs) measured in elastic electron scattering provide information on the transverse charge and current distributions inside the nucleon~\cite{Hofstadter:1955}. On the other hand, parton 
distribution functions (PDFs) measured in Deeply Inelastic Scattering (DIS) characterize the longitudinal momentum distribution of the underlying quarks and gluons~\cite{Friedman:1991}. Twenty years ago, FFs and PDFs were unified within the formalism of Generalized Parton Distributions (GPDs)~\cite{Mueller:1998fv,Ji:1996ek,Radyushkin:1997ki}. GPDs are universal functions encoding a wealth of information about the nucleon internal structure such as the correlation between the transverse position of quarks and gluons  (partons)
and their longitudinal momenta~\cite{Burkardt:2003}. GPDs also provide access to the contribution of quark and gluon orbital angular momenta to the nucleon spin~\cite{Ji:1996ek}. Eight GPDs for each quark flavor $q$ describe nucleon structure at leading order in $1/Q$ (twist-2). They correspond to each combination of nucleon and parton helicities. The four chiral-even GPDs ($H^q$, $E^q$, $\widetilde{H}^q$ and $\widetilde{E}^q$) conserve the helicity of the parton whereas the four chiral-odd, or  transversity GPDs ($H_T^q$, $E_T^q$, $\widetilde{H}_T^q$ and $\widetilde{E}_T^q$), flip the parton helicity~\cite{Diehl:2003,Hoodbhoy:1998vm}.

GPDs parametrize the structure of the target independently
of the reaction. Chiral-even GPDs can be accessed experimentally via hard exclusive processes such as deeply virtual Compton scattering (DVCS) and deeply virtual meson 
electroproduction (DVMP) in the Bjorken limit $Q^2\to\infty$ and $t/Q^2\ll 1$ at fixed $x_B$. Recent results on DVCS show the validity of this limit at values of $Q^2$ as low as 1.5~GeV$^2$~\cite{MunozCamacho:2006hx,Defurne:2015kxq,Jo:2015ema}. 
In the case of DVMP, the longitudinal scattering amplitude factorizes into a hard perturbative contribution and a soft convolution of the nucleon GPDs and the meson distribution amplitude (DA). The transverse virtual photo-production amplitude is proven to be suppressed by a factor of $1/Q^2$ at sufficiently high values of $Q^2$~\cite{Collins:1996fb}. In the case of $\pi^0$ electroproduction, it was suggested in~\cite{Ahmad:2008hp, Goloskokov:2011rd} that a large contribution to the transverse amplitude could arise from the convolution of the transversity GPDs of the nucleon with a twist-3 quark-helicity flip pion DA. Model calculations including the transversity GPDs have successfully described recent  $\pi^0$ electroproduction data on a proton  target, measured at Jefferson Lab (JLab)~\cite{Collaboration:2010kna,Bedlinskiy:2012be,Bedlinskiy:2014tvi,Defurne:2016}.
Measurements of $\pi^0$ electroproduction on the neutron are extremely interesting as they provide the exciting possibility to separate the individual contributions of the $u$ and $d$ quarks to the cross sections, when combined with measurements from a proton target at the same kinematics. 

 \begin{figure}
\begin{minipage}{0.65\linewidth}
    \includegraphics[width=0.9\linewidth]{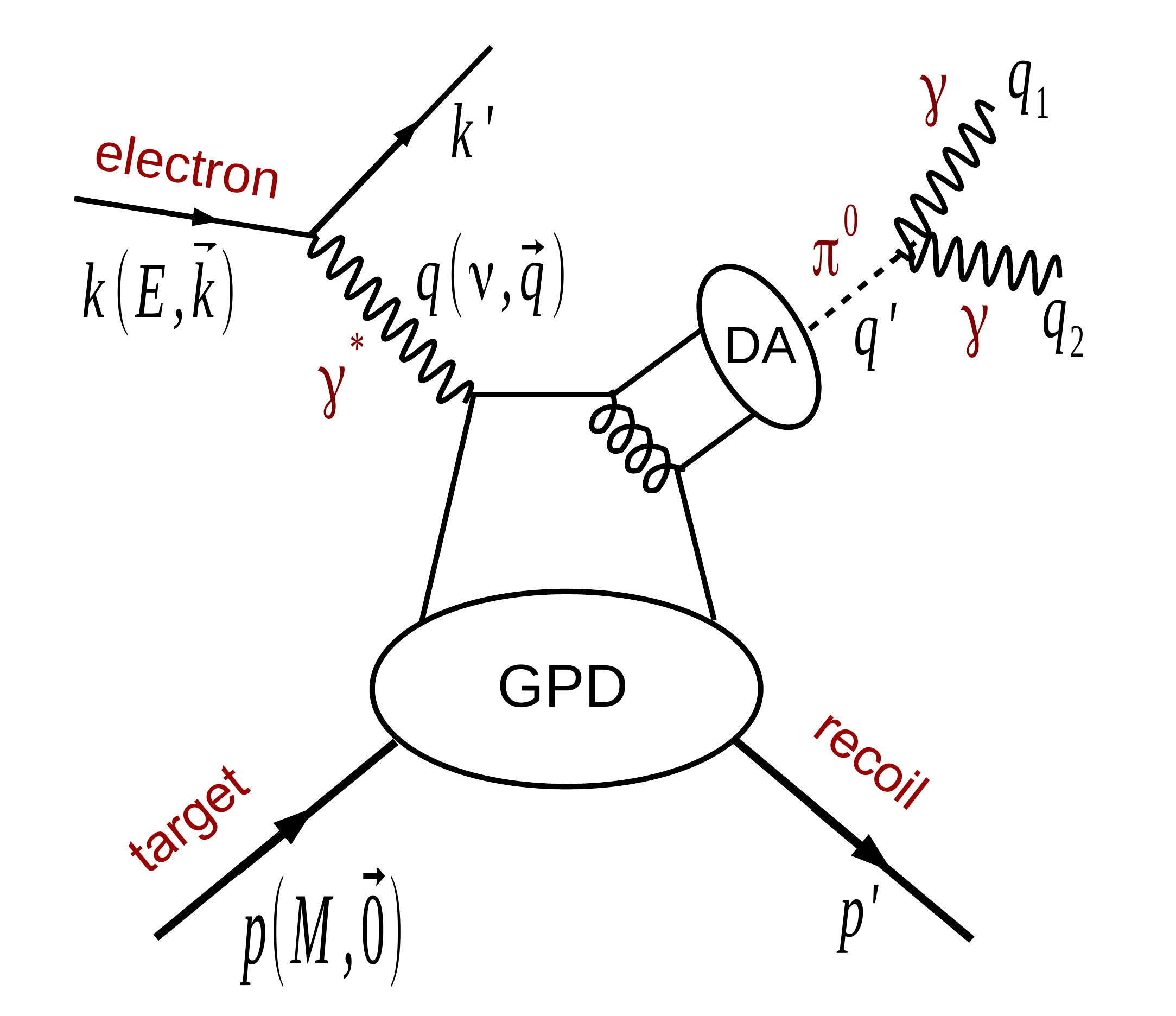}
    \end{minipage}\hfill\begin{minipage}{0.34\linewidth}
    \centerline{Invariants}
    \vskip -1.5em
    \begin{align*}
    Q^2 &= -(k-k')^2 \\
    x_A &= Q^2/(2q\cdot p) \\
    W^2 &= (q+p)^2 \\
    y    &= (q\cdot p)/(k\cdot p)\\
    t      &= (q-q')^2  \\
    t^{\prime}      &= t_\text{min}-t  \\
    \end{align*}
    \end{minipage}
    \caption{Diagram of the coherent $\pi^0$ electroproduction reaction on the nucleon ($M=M_N$, $x_A = x_B$) or deuteron ($M=M_d$, $x_A=x_d$) with the dominant $\pi^0\to \gamma\gamma$ decay mode. 
    The minimal $|t|$ value is 
    $t_\text{min} = (Q^2+m_{\pi}^2)^2/(4W^2)-(|\vec{q}^{~ c.m.}|-|\vec{q}^{~\prime c.m.}|)^2$, where $m_{\pi}$ is the $\pi^0$ mass and the $c.m.$ superscript refers to the
     target$-\pi^0$ center-of-mass frame.}
    \label{figure1}
  \end{figure}

The differential cross section of deeply virtual $\pi^0$ production is given by~\cite{DT}:
\begin{multline}
\frac{d^4\sigma}{dQ^2 dx_A dt d\phi}=\frac{1}{2\pi}\frac{d^2\Gamma_A}{dQ^2 dx_A}
\Big[\frac{d\sigma_T}{dt}+\epsilon\frac{d\sigma_L}{dt}\\
\sqrt{2\epsilon (1+\epsilon)}\frac{d\sigma_{TL}}{dt}\cos\phi+\epsilon \frac{d\sigma_{TT}}{dt}\cos 2\phi\Big]\;,
\label{sigtot}
\end{multline}
where  $\phi$ is  the angle between the hadronic and leptonic planes following the Trento Convention~\cite{Bacchetta:2004jz}. The virtual photon flux factor $d^2\Gamma_A$ and photon polarization $\epsilon$ are  defined by:
\begin{align}
\frac{d^2\Gamma_A}{dQ^2 dx_A} &= \frac{\alpha}{2\pi} \frac{y^2(1-x_A)}{x_A Q^2}\frac{1}{1-\epsilon}~, \nonumber \\
\epsilon &= \frac{1-y- Q^2/(2E)^2}{1-y+y^2/2 + Q^2/(2E)^2}~.
\end{align}
Fig.~\ref{figure1} shows the lowest order Feynman diagram of the reaction and includes definitions of the kinematic variables.
The $\phi$ dependence in Eq.~(\ref{sigtot}) allows the extraction of the interference terms $d\sigma_{TL}/dt$ and $d\sigma_{TT}/dt$ while  measurements of the total cross section at two incident beam energies and fixed $Q^2$ and $x_B$ separate $d\sigma_T/dt$  and $d\sigma_L/dt$.

\begin{figure}[b]
\centering 
\includegraphics[width=0.9\linewidth]{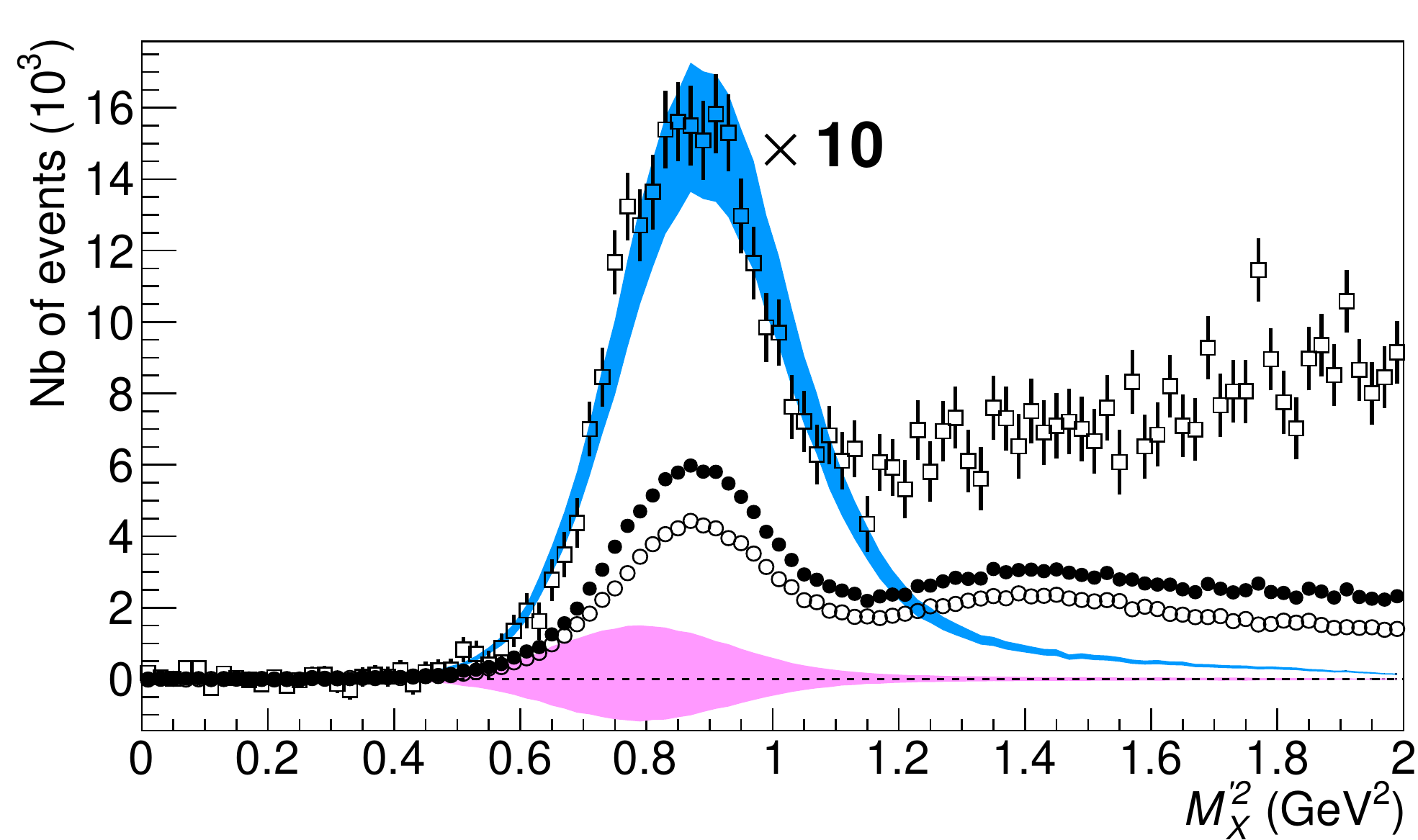}
\caption{Corrected missing mass squared $M_X^{\prime 2}$ for D$(e,e'\pi^0)X$ (solid circles) and 
normalized Fermi-smeared H$(e,e'\pi^0)X$  events (open circles). Bars show statistical uncertainties. The difference between the two distributions (squares) is scaled by a factor 10 for clarity. The blue and magenta bands (both scaled $\times 10$), show the simulated $n(e,e^{\prime}\pi^0)n$  and $d(e,e^{\prime}\pi^0)d$ yields, 
respectively,  fit to the data by minimizing Eq.~(\ref{eq::fit}).  These bands  include the statistical uncertainty of the fit.}
\label{figure2}
\end{figure}

In JLab Hall A experiment E08-025, we measured the D$(e,e'\pi^0)X$ reaction, with the primary goal of extracting the $n(e,e\pi^0)n$ cross
section in the quasi-free approximation. We perform a Rosenbluth separation, based on data taken with incident beam energies $E= 4.455$ and
5.550 GeV.   A 15-cm-long liquid deuterium (LD2) target was used as a quasi-free neutron target. The quasi-free
$\pi^0$ electroproduction events off the proton are subtracted using the data from experiment E07-007~\cite{Defurne:2016}. These two experiments ran concurrently with  liquid hydrogen (LH2)  and LD2 targets  interchanged daily  to minimize systematic uncertainties.
 Scattered electrons were detected in the left High Resolution Spectrometer (HRS) of Hall A~\cite{Alcorn:2004sb}, which determined accurately the electron scattering kinematics centered at $x_B =0.36$ and $Q^2=1.75\text{ GeV}^2$. The two photons from the $\pi^0$ decay were detected in an electromagnetic calorimeter composed of a $13\times 16$  array of $3\times3\times 18.6\text{ cm}^3$ PbF$_2$ crystals, resulting in a 
$[0,2\pi]$ coverage in $\phi$ and $[0,0.25]$ GeV$^2$ range in $t^{\prime}=t_{min}-t$. A 0.6~ns $\pi^0$-electron coincidence time resolution was achieved by means of a 1~GHz 
flash ADC system in each calorimeter channel. The calibration of the calorimeter was performed with elastic 
H($e$,$e_\text{Calo}^\prime p_\text{HRS}$) data from 
dedicated runs in which the  scattered electrons  were detected in the calorimeter, with energy predetermined by the kinematics
of the elastic recoil proton in the HRS.
The calorimeter calibration was monitored continuously \textit{ a postiori} by tracking the 
2-photon invariant mass $m_{\gamma\gamma}=\sqrt{(q_1+q_2)^2}$ and the $ep\to e\pi^0X$ missing mass squared $M_X^2=(q+p-q_1-q_2)^2$.
Exclusive $\pi^0$ electroproduction events are  selected for each $(t^{\prime},\phi)$ bin by applying a bidimensional cut:
\begin{eqnarray}
\left|m_{\gamma\gamma}- m_{\pi}\right| &<& 4~\sigma_{m_{\gamma\gamma}} ~, \\    
M_X^{\prime 2}=M_X^2+C~(m_{\gamma\gamma}-m_{\pi}) ~~&<& ~0.95 ~\text{GeV}^2~,
\label{cut}
\end{eqnarray}
where $\sigma_{m_{\gamma\gamma}}$ is the resolution of the reconstructed $\pi^0$ invariant mass, and the empirical factor $C=13$~GeV takes into account the natural correlation between the invariant mass and missing mass originating from energy fluctuations in the calorimeter. 
Fig.~\ref{figure2} shows the corrected missing mass squared $M_X^{\prime 2}$ obtained at $E$=4.455 GeV for LH2 and LD2 data sets where $M_X^2$ is calculated with a target corresponding to a nucleon at rest. Accidentals were subtracted from these spectra and the LH2 data were normalized to the same 
integrated luminosity as the LD2 data.

The average momentum transfer to the target $\langle|\vec{\Delta}|\rangle=\langle|\vec{q}-\vec{q^{\prime}}|\rangle$ in the
kinematics of this experiment is much larger than 
the average $np$ relative momentum in the deuteron wavefunction $\langle|\vec{p_F}|\rangle$ . Below the threshold for the production of a second pion, 
the impulse approximation is expected to accurately describe the exclusive 
D$(e,e^{\prime}\pi^0)X$  yield, with $X = np \oplus d$.  Thus we write the cross section as the sum
 of the  coherent elastic channel $d(e,e^{\prime}\pi^0)d$ and two incoherent quasi-elastic contributions: 
\begin{equation}
D(e,e^{\prime}\pi^0)X = d(e,e^{\prime}\pi^0)d + n(e,e^{\prime}\pi^0)n + p(e,e^{\prime}\pi^0)p.
\label{impulse}
\end{equation}
We subtract the $ p(e,e^{\prime}\pi^0)p$ yield from the deuterium data by normalizing our H$(e,e^{\prime}\pi^0)X$ data
to the luminosity of the LD2 data.  The  Fermi-momentum $\vec{p_F}$ of bound protons inside the deuteron is statistically added to the LH2 data following the distribution given in~\cite{Lacombe:1980} since this effect is intrinsically present in the $M_X^{\prime 2}$ spectrum of the LD2 data.
The result of the subtraction of the H$(e,e^{\prime}\pi^0)X$ data from the D$(e,e^{\prime}\pi^0)X$ yield is shown in Fig.~\ref{figure2}. 
The $d(e,e^{\prime}\pi^0)d$  and $n(e,e^{\prime}\pi^0)n$ channels are \textit{in-principle} kinematically separated by
$\Delta M_X^{\prime 2} =t(1-M/M_d)\approx t/2$ where $M_d$ is the deuteron mass. This kinematic shift is exploited in the procedure described below to separate the contributions of the quasi-free neutron and coherent deuteron channels in the total $\pi^0$ electroproduction cross section. 

Fig.~\ref{figure2} illustrates that the exclusive $\pi^0$ electroproduction events are primarily localized below the  production threshold for
a second pion:  $M_X^{\prime 2} < (M+m_{\pi})^2\approx 1.15\text{ GeV}^2$. 
However, we apply a  nominal cut of  $M_X^{\prime 2} <0.95\text{ GeV}^2$  to minimize any contamination of inclusive events that
might arise from  resolution effects. The resulting events below this  $M_X^{\prime 2}$ cut are divided into $12\times 2 \times 5 \times 30$ bins in $\phi$, $E$ , $t^{\prime}$ and $M_X^{\prime 2}$ respectively. 
The first two variables allow the independent extraction of the four structure functions of the $\pi^0$ electroproduction cross section while the binning in $M_X^{\prime 2}$ enables the separation of the  $d(e,e^{\prime}\pi^0)d$ and $n(e,e^{\prime}\pi^0)n$ contributions.

A Monte-Carlo simulation of the experimental setup is based on the {\sc Geant4} toolkit~\cite{Agostinelli:2003}. It includes both external and real internal radiative effects based on calculations described in 
~\cite{Vanderhaeghen:2000ws}. The virtual internal effects are applied as a global correction factor to the extracted cross sections. The HRS acceptance is modeled by an R-function~\cite{Rvachev:2001} defining correlated multi-dimensional boundaries. Only the overlapping ($Q^2$,$x_B$) phase-space between the two beam energy settings is considered.
The calorimeter energy resolution in the $p(e,e^{\prime}\pi^0)p$ simulation is smeared to match the $M_X^{\prime 2}$ distribution in each ($E$, $t^{\prime}$, $\phi$) bin of the LH2 data.  These bin-by-bin resolution
smearing factors are also applied to the $n(e,e^{\prime}\pi^0)n$ and $d(e,e^{\prime}\pi^0)d$ simulated data. The Fermi-smearing
described above is also applied to the simulated $n(e,e^{\prime}\pi^0)n$ yields. The systematic uncertainty of this smearing procedure as well the asymmetric systematic uncertainty originated from the inclusive yield under the $M_X^{\prime 2}$ cut are evaluated by varying the cut applied around its nominal value. They are found to be bin-dependent and were added quadratically to the $3.1\%$ normalization uncertainty listed in~\cite{Defurne:2016}.

For each $t^{\prime}$ we fit the simulated yield to the experimental distributions of the   $\phi$- and $M_X^{\prime 2}<\text{cut}$-bins.
To wit, we minimize the  $\chi^2$:
\begin{equation}
\chi^2=\sum_{i=1}^{3600}\left(\frac{N^{exp}_i-N^{sim}_i}{\delta^{exp}_i}\right)^2\;,
\label{eq::fit}
\end{equation}
where $N^{exp}_i$ ($N^{sim}_i$) is the number of experimental (simulated) events in bin $i$ and $\delta^{exp}_i$ is the corresponding uncertainty. The kinematic factors appearing in Eq.~(\ref{sigtot}) are convoluted with the experimental acceptance and resolution
in the computation of $N^{sim}_i$. The eight cross-section structure functions $d\sigma_\Lambda^{n,d}(t')/dt$ ($\Lambda = T, L, LT, TT$)
which define $N^{sim}_i$ are the free parameters of the fit for each $t^{\prime}$ bin.

\begin{figure}[]
\centering
\includegraphics[width=\linewidth]{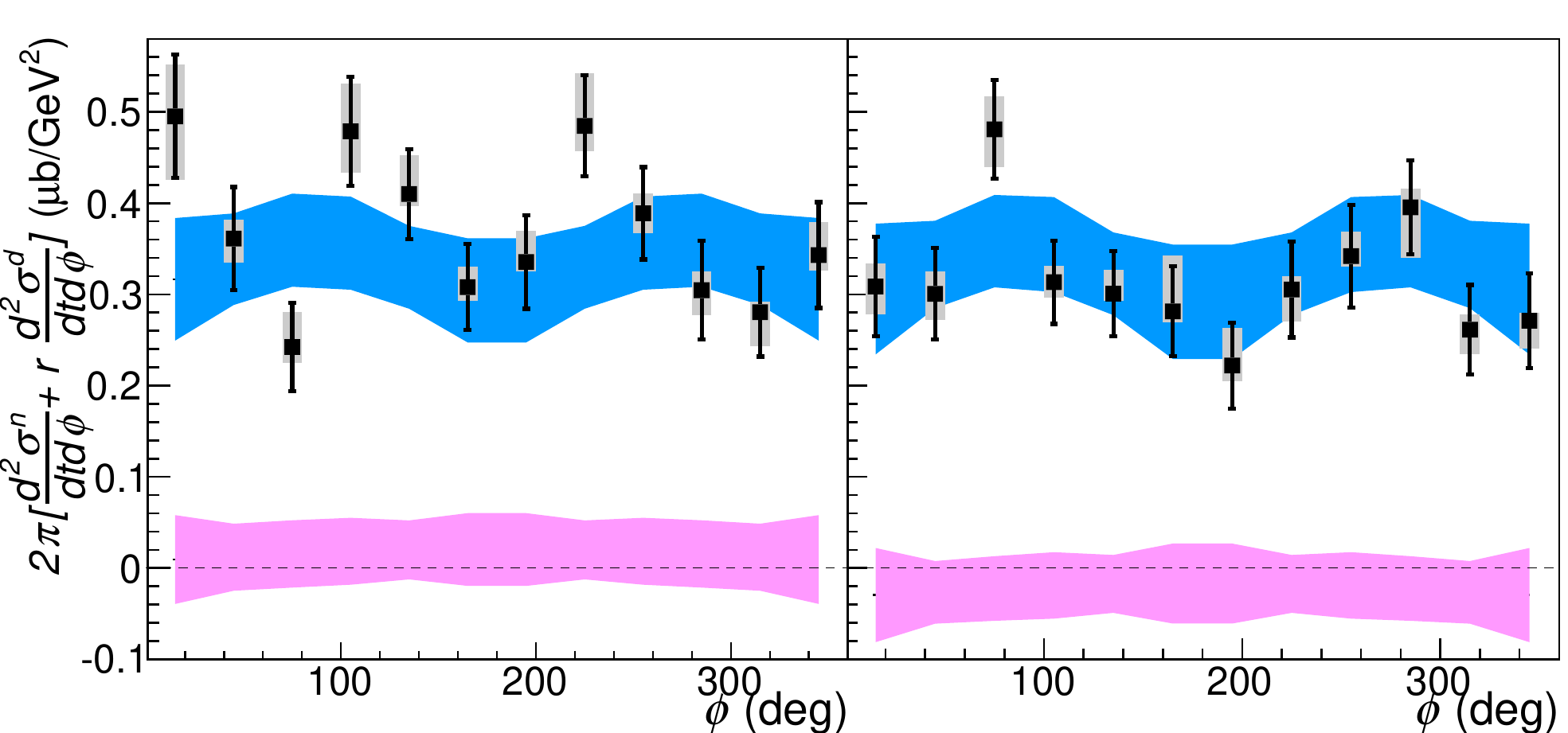}
\caption{Total cross section 
$2\pi\left(\frac{d^2\sigma^n}{dt d\phi}+ r \frac{d^2\sigma^d}{dtd\phi}\right)$
 as a function of $\phi$ 
at $E =4.45$ GeV (left) and $E=5.55$ GeV (right), in the bin $\left<t^{\prime}\right>=0.025\text{ GeV}^2$ (neutron kinematics),
equivalently $\left<t^{\prime}\right>$=0.021 GeV$^2$ (deuteron kinematics), with r=1.27 (left) and r=1.33 (right) being the ratio deuteron$/$neutron of the virtual photon flux convoluted with the experimental acceptance.
The error-bars show the statistical uncertainty. 
Filled grey boxes around the points show the total systematic uncertainties.
The blue and magenta bands represent the contributions of $2\pi\frac{d^2\sigma^n}{dt d\phi}$ and
 $2\pi\frac{d^2\sigma^d}{dt d\phi}$, respectively, including the statistical uncertainty of the fit.}
\label{figure3}
\end{figure}

\begin{figure}[]
\centering
\includegraphics[width=\linewidth]{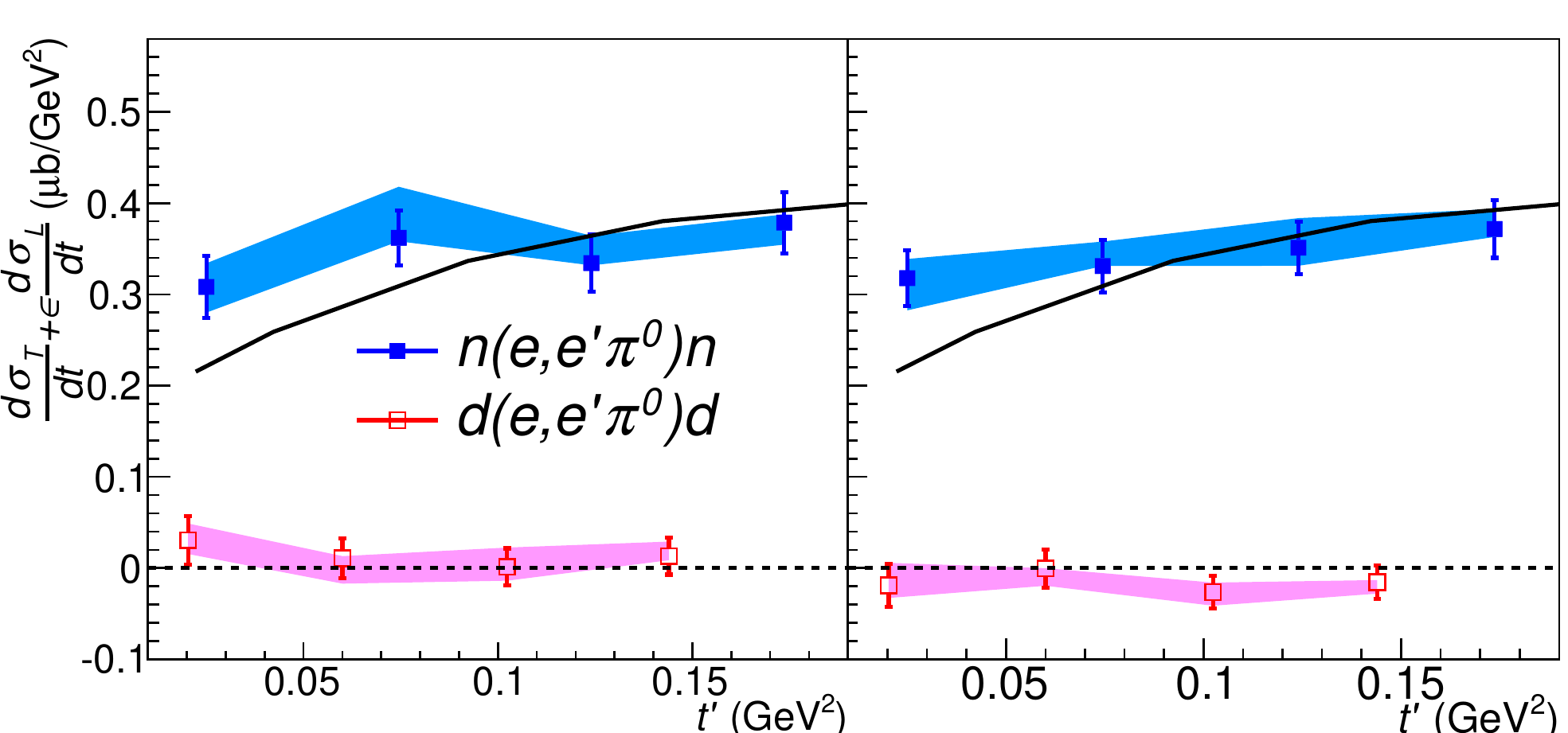}
\caption{The $\phi$-independent photo-production cross sections extracted from the fit, as functions of $t'$,
and separated into quasi-free neutron
and coherent deuteron contributions:
$\frac{d\sigma_T^n}{dt}+ \epsilon \frac{d\sigma_L^n}{dt}$ and
$\frac{d\sigma_T^d}{dt}+ \epsilon \frac{d\sigma_L^d}{dt}$.
The data in the left and right panels were obtained 
at $E=4.45$ GeV and  $E=5.55$  GeV, respectively.  The error-bars show the statistical uncertainty from the fit.
The blue and magenta bands represent the systematic errors. 
The solid lines are theoretical calculations for the neutron from~\cite{Goloskokov:2011rd}. 
}
\label{figure4}
\end{figure}

\begin{figure}[]
\centering 
\includegraphics[width=0.9\linewidth]{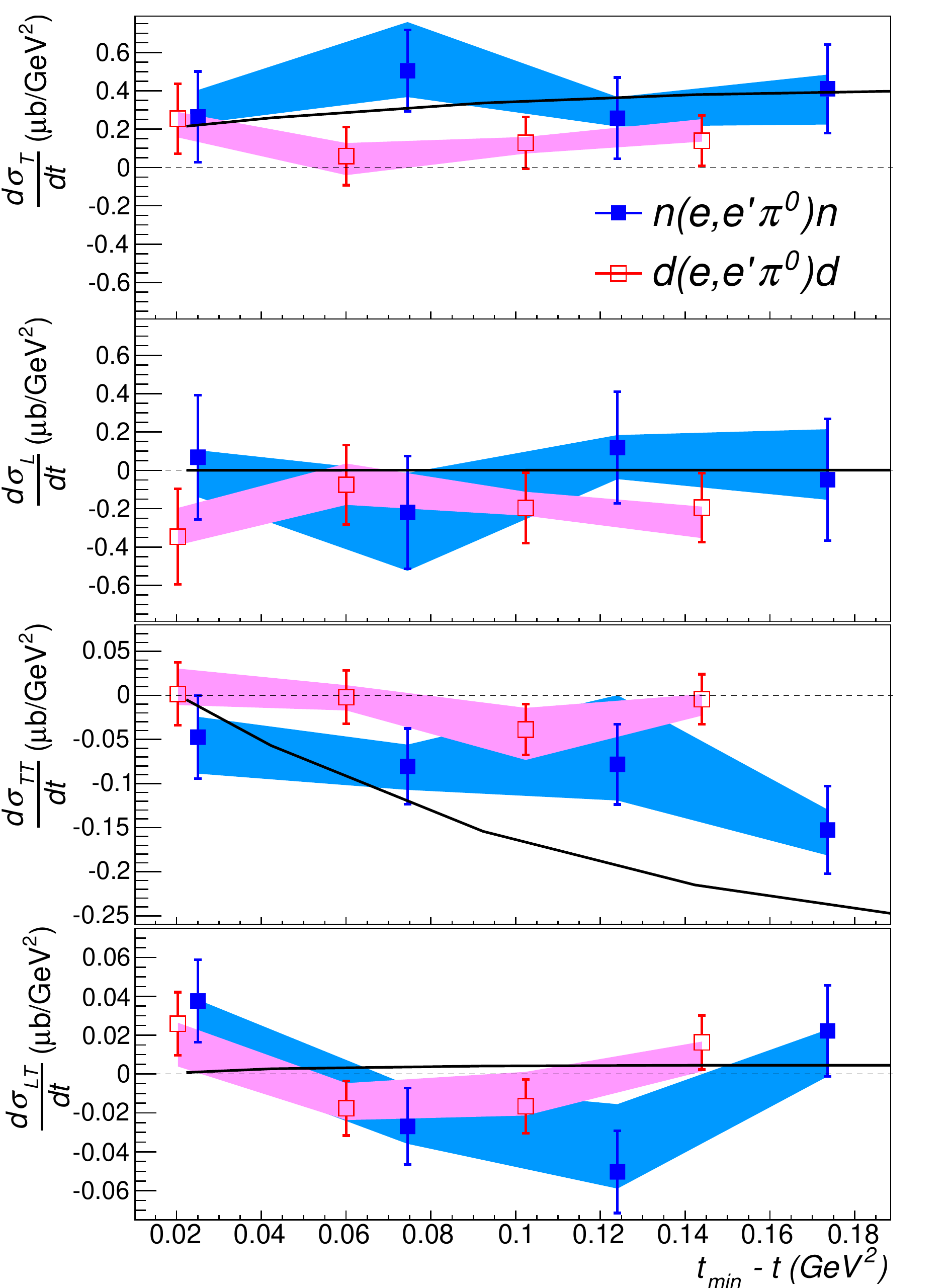}
\caption{Structure functions $d\sigma_T/dt$, $d\sigma_L/dt$, $d\sigma_{TL}/dt$ and $d\sigma_{TT}/dt$ as a function of $t^{\prime}=t_{min}-t$ for the neutron (blue) and the deuteron (red). The filled bands around the points show systematic uncertainties. The solid lines are theoretical calculations for the neutron from~\cite{Goloskokov:2011rd}. }
\label{figure5}
\end{figure} 

Fig.~\ref{figure3} shows the measured $\phi$-dependent  photo-absorption cross section 
for both beam energies and for the lowest $t^{\prime}$ bin. The $d^2\sigma^n/{dtd\phi}$ cross section 
is almost independent of the beam energy indicating a dominance of the transverse response. 
The  ${d^2\sigma^d}/{dtd\phi}$ cross section is found negligible within uncertainties for all $\phi$ bins. 
The fit to the $M_X^{\prime 2}$-distribution is  shown in Fig.~\ref{figure2} which also illustrates that  the LD2--LH2 yield is dominated by the neutron contribution in the exclusive region.
In Fig.~\ref{figure4}, we display $\phi$-independent cross section $d\sigma_T + \epsilon d\sigma_L$ for the two beam energies,
separated into the fitted quasi-free neutron and coherent deuteron channels. The highest $t^{\prime}$ bin
is used in the analysis to treat bin migration effects and is not shown herein. The figure again shows the clear separation of the neutron
signal. The coherent deuteron cross sections are found to be very small and compatible with theoretical calculations based on chiral-even deuteron GPDs, which  predict cross-section values smaller than 1 nb/GeV$^2$ in similar kinematics~\cite{Pire:2004epj}.

Fig.~\ref{figure5} shows the four extracted structure functions for the neutron and the deuteron as functions of $t^{\prime}$.  The neutron cross sections are dominated by $d\sigma_T^n/dt$ and $d\sigma_{TT}^n/dt$, while the terms involving a longitudinal response are compatible with zero within uncertainties and are in good agreement with previous results off a proton target at the same kinematics~\cite{Defurne:2016}.  
The neutron measurements are compared to a calculation based on both quark helicity-conserving GPDs and quark helicity-flip
(transversity) GPDs~\cite{Goloskokov:2011rd}, and show  good agreement for all structure functions, with a slight overestimation of  
$|d\sigma_{TT}^n/dt|$. 
The  experimental $d\sigma_{L}^n/dt$ term is also compatible with the VGG model~\cite{Vanderhaeghen:1999xj} based on chiral-{even} 
 GPDs, which  predicts $d\sigma_{L}^n/dt < 4\text{ nb/GeV}^2$ for all $t^{\prime}$ bins. 
 
 Together with previous measurements of $d\sigma_T/dt$ and $d\sigma_{TT}/dt$ on the proton~\cite{Defurne:2016} and extensive unseparated measurements before~\cite{Collaboration:2010kna,Bedlinskiy:2012be,Bedlinskiy:2014tvi}, these new results provide strong support to the exciting idea that transversity GPDs can be accessed via  neutral pion electroproduction in the
high $Q^2$ regime.

Within the modified factorization approach of~\cite{Goloskokov:2011rd}, 
$d\sigma_T/dt$ and $d\sigma_{TT}/dt$ are functions of $\langle H_T\rangle$ and $\langle\bar{E}_T\rangle$, which are convolutions of the 
elementary $\gamma^* q \to q^{\prime} \pi^0$ amplitude with the transversity GPDs 
$H_T$ and $\bar{E}_T=2\widetilde{H}_T+E_T$:
\begin{eqnarray}
\frac{d\sigma_T}{dt} &=& \Lambda ~\left[ \left( 1-\xi^2 \right) \left|\left<H_T\right>\right|^2 -\frac{t^{\prime}}{8M^2}\left|\left<\bar{E}_T\right>\right|^2\right]\,,\\
\frac{d\sigma_{TT}}{dt} &=& \Lambda ~\frac{t^{\prime}}{8M^2}\left|\left<\bar{E}_T\right>\right|^2 \,.
\label{sigtt}
\end{eqnarray}

\begin{figure}[b!]
\centering
\includegraphics[width=0.9\linewidth]{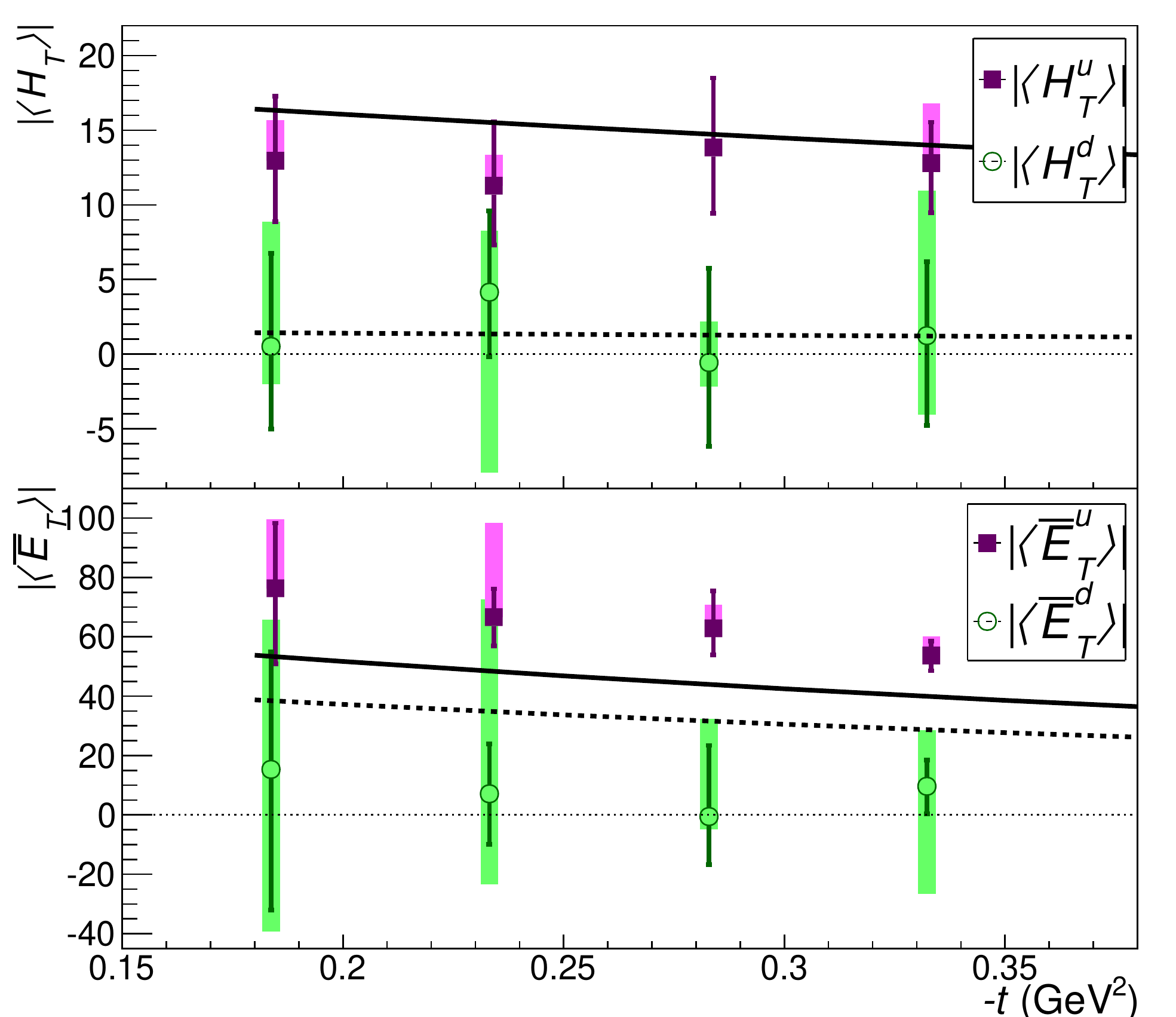}
\caption{Magnitude of the nucleon helicity-flip $\left<H_T\right>$ (top) and non-flip $\left<\bar{E}_T\right>$ (bottom) 
transversity terms for $u$ (squares) and $d$ (circles) quarks assuming no relative phase between them. 
Filled boxes around the points represent the variation of the results when their relative phase varies between 0 and $\pi$. Bars show the quadratic sum of the statistical and systematic uncertainties of the data. Solid (dashed) lines are calculations from the Goloskokov-Kroll model~\cite{Goloskokov:2011rd} for $u$ ($d$) quark.}
\label{figure6}
\end{figure}

In these equations $\Lambda(Q^2,x_B)$ is a phase space factor~\cite{Bedlinskiy:2014tvi} and $\xi\simeq x_B/(2-x_B)$ is the skewness variable. 
For a proton and a neutron target, the quark-flavor structures of  $\left|\left<H_T\right>\right|^2$ (neglecting strange quarks) are:

\begin{align}
\left|\left<H_T^{p,n}\right>\right|^2 &= \frac{1}{2} \left|\frac{2}{3}\left<H_T^{u,d}\right> + \frac{1}{3}\left<H_T^{d,u}\right> \right|^2,
\label{GPD}
\end{align}
with similar equations for $\left|\left<\bar{E}_T\right>\right|^2$.
The different flavor weights of the proton and neutron targets allow us to separately determine $\left|\left<H_T^u\right>\right|$ and $\left|\left<H_T^d\right>\right|$ (similarly $\left|\left<\bar{E}_T^u\right>\right|$ and $\left|\left<\bar{E}_T^d\right>\right|$) by combining the data we report herein and $\pi^0$ electroproduction cross sections on the proton measured at the same kinematics as in~\cite{Defurne:2016}. The unknown relative phase between the $u$ and $d$ convolutions is treated as a systematic uncertainty in the separation. The flavor-separated results assuming no relative phase between the $u$ and $d$ convolutions are presented in Fig.~\ref{figure6}, with the bands indicating their variation when the phase takes all possible values between 0 and $\pi$. This phase could be resolved with  exclusive $p(\gamma^\ast, \eta p)$ data in the same kinematics.
Fig.~\ref{figure6} shows that the magnitudes of the $u$-quark convolutions are  larger than the $d$-quark convolutions for all $t$ bins. 
The results in Fig.~\ref{figure6} also demonstrate that the $u$-quark nucleon helicity non-flip term
 $\left|\left<\bar{E}_T^u\right>\right|$, 
 is larger than the nucleon helicity flip term $\left|\left<H_T^u\right>\right|$.
The comparison to the Goloskokov-Kroll model~\cite{Goloskokov:2011rd} shows  good agreement for $\left|\left<H_T\right>\right|$ for both quark flavors but an underestimation for $\left|\left<\bar{E}_T^u\right>\right|$.  
The GPD $H_T$ parametrization is constrained in the forward limit by the transversity parton distributions. However, no similar experimental constraint is available for $\bar{E}_T$.  The constraints on $\bar{E}_T$ are mainly taken from lattice QCD calculations~\cite{Gockeler:2007}.  

In conclusion, we have separated the four unpolarized structure functions of $\pi^0$ electroproduction off the neutron at $Q^2$=1.75 GeV$^2$ and $x_B$=0.36 in 
the $t^{\prime}$ range $[0,0.2]$ GeV$^2$. Similar measurements are obtained for coherent $\pi^0$ electroproduction off the deuteron at $x_d$=0.18. The latter are found to be very small and according to theoretical expectations. Neutron results show a dominance of the transverse response confirming the transversity GPD approach for the description of this process. By combining neutron and proton results, we have performed the first flavor decomposition of the $u$ and $d$ quark contributions to the cross section. 

We thank P.~Kroll, S.~Goloskokov, M.~Guidal, M.~Vanderhaeghen and B.~Pire for valuable information about their work and providing the results of their models. We acknowledge essential work of the JLab accelerator staff and the Hall A technical staff. This work was supported by the Department of Energy (DOE), the National Science Foundation, the French {\em Centre National de la Recherche Scientifique}, the {\em Agence Nationale de la Recherche}, the {\em Commissariat \`a l'\'energie atomique et aux \'energies alternatives} and P2IO Laboratory of Excellence. Jefferson Science Associates, LLC, operates Jefferson Lab for the U.S. DOE under U.S. DOE contract DE-AC05-060R23177.

\bibliography{Compton2014}

\end{document}